\documentclass{article}

\usepackage{microtype}
\usepackage{graphicx}
\usepackage{booktabs}
\usepackage{hyperref}

\usepackage{xcolor}

\usepackage{arxiv}

\makeatletter
\renewcommand{\shorttitle}{Epistemic Status and Temporal Validity in AI-Assisted Engineering}
\makeatother

\usepackage{natbib}

\usepackage{amsmath}
\usepackage{amssymb}
\usepackage{amsthm}
\newtheorem{theorem}{Theorem}

\usepackage{tikz}
\usetikzlibrary{arrows.meta, positioning, shapes.geometric, calc,
                 decorations.pathreplacing, patterns, fit}
\usepackage{pgfplots}
\pgfplotsset{compat=1.18}

\definecolor{fpfblue}{RGB}{70,130,180}
\definecolor{fpfgray}{RGB}{120,120,130}
\definecolor{fpfamber}{RGB}{200,150,50}
\definecolor{fpfgreen}{RGB}{80,160,80}
\definecolor{fpfred}{RGB}{190,70,70}
\definecolor{fpflightblue}{RGB}{210,230,250}
\definecolor{fpflightgreen}{RGB}{215,240,215}
\definecolor{fpflightamber}{RGB}{255,240,210}
\definecolor{fpflightred}{RGB}{250,220,220}

\title{AI-Assisted Engineering Should Track the Epistemic Status and Temporal Validity of Architectural Decisions}

\author{
  Sankalp Gilda$^{*}$ \\
  DeepThought Solutions \\
  \texttt{sankalp@deepthoughtsolutions.xyz} \\
  \texttt{sankalp.gilda@gmail.com}
  \And
  Shlok Gilda$^{*}$ \\
  Department of Computer Science \\
  University of Florida \\
  \texttt{shlokgilda@ufl.edu}
}

\date{}

\begin{document}

\maketitle

\renewcommand{\thefootnote}{\fnsymbol{footnote}}
\footnotetext[1]{Equal contribution.}
\renewcommand{\thefootnote}{\arabic{footnote}}

% ===========================================================================
% ABSTRACT
% ===========================================================================
\begin{abstract}
This position paper argues that AI-assisted software engineering requires
explicit mechanisms for tracking the epistemic status and temporal validity of
architectural decisions. LLM coding assistants generate decisions faster than
teams can validate them, yet no widely-adopted framework distinguishes
conjecture from verified knowledge, prevents trust inflation through
conservative aggregation, or detects when evidence expires. We propose three
requirements for responsible AI-assisted engineering: (1)~epistemic layers that
separate unverified hypotheses from empirically validated claims,
(2)~conservative assurance aggregation grounded in the G\"{o}del t-norm that
prevents weak evidence from inflating confidence, and (3)~automated evidence
decay tracking that surfaces stale assumptions before they cause failures. We
formalize these requirements as the First Principles Framework (FPF), ground
its aggregation semantics in fuzzy logic, and define a quintet of invariants
that any valid aggregation operator must satisfy. Our retrospective audit
applying FPF criteria to two internal projects found that 20--25\% of architectural decisions had
stale evidence within two months, validating the need for temporal
accountability. We outline research directions including learnable aggregation
operators, federated evidence sharing, and SMT-based claim validation.
\end{abstract}


% ===========================================================================
% 1. INTRODUCTION
% ===========================================================================
\section{Introduction}\label{sec:introduction}

Modern LLM coding assistants (GitHub Copilot, Cursor, Claude Code, Gemini Code
Assist) speed up software development but open a gap: decisions are made faster
than they can be validated. A developer asks ``Should I use Redis or
Memcached?'' The AI gives an answer that lacks epistemic qualification, the
developer implements it immediately, and nobody revisits the assumptions when a
library update or infrastructure change invalidates the original rationale.
Software engineering is the canary for a broader problem. In drug interaction
screening, a model like Med-PaLM~\citep{singhal2023medpalm} may assess a
combination as safe based on training data that predates an FDA
contraindication update, and nothing in the system tracks that the underlying
evidence has expired. In legal contract analysis, an AI assistant drafts
clauses referencing case law that was subsequently overturned; the output
carries no metadata distinguishing established precedent from dicta in an
unpublished opinion. In autonomous scientific
experimentation~\citep{boiko2023autonomous}, a lab agent selects reagent
concentrations from literature values obtained under different assay
conditions, with no distinction between ``validated in our setup'' and
``worked in a related context.'' ML engineering suffers the same problem
reflexively: models are selected based on benchmark results that predate
evaluation-harness changes, hyperparameter choices cite ablation studies from
different data distributions, and deployment decisions assume training-time
metrics transfer to production, each decision cached as institutional
knowledge and rarely revisited.

We ground the paper in software engineering because it has the most deployment
data available today, but the framework applies to any domain where decisions
rest on evidence that expires.

Four problems follow:

\begin{enumerate}
\item No distinction between ``untested hypothesis'' and ``empirically verified
  claim.'' A cached LLM suggestion and a load-tested benchmark carry equal
  weight in team memory.
\item Multiple weak arguments get averaged into strong-seeming evidence. Three
  blog posts do not equal one controlled experiment, but informal reasoning
  treats them as comparable.
\item No tracking of when evidence expires. Benchmarks go stale, dependencies
  update, requirements change, yet decisions persist as if the world were
  frozen.
\item No audit trail showing why decisions were made or what conditions would
  invalidate them. Post-mortems repeatedly find ``nobody remembers why we
  chose~X.''
\end{enumerate}

A recent survey of 47 academic studies on generative AI for software
architecture~\citep{esposito2025genai} finds that 93\% of surveyed papers report no validation of
LLM-generated architectural outputs. Current LLM reasoning approaches address
parts of this problem. Self-consistency
voting~\citep{wang2023selfconsistency} aggregates multiple reasoning paths.
Verifier scoring~\citep{lightman2024verify} rates individual reasoning steps.
Chain-of-thought prompting~\citep{wei2022chainofthought} produces
interpretable traces. But none of these approaches provide explicit epistemic
layers, temporal validity tracking, conservative aggregation with formal
guarantees, or durable audit trails that survive beyond a single session.

\textbf{Our position:} AI-assisted engineering workflows demand
\emph{mathematical guarantees against epistemic drift}. We formalize this
through a principled framework with three core properties: First, explicit
epistemic layers must distinguish unverified hypotheses~(L0) from empirically
validated claims~(L2). Second, conservative assurance aggregation requires
adherence to a \emph{quintet of mathematical invariants}, ensuring that no
conclusion can be more reliable than its weakest supporting evidence (the WLNK
bound), and that formality levels impose hard reliability ceilings regardless
of consensus: ten informal observations cannot equal one controlled experiment.
Third, evidence validity windows must be tracked mechanically through automated
alerts, not through aspirational review schedules that teams deprioritize under
delivery pressure. These claims are falsifiable: a practitioner could reasonably
argue that averaging better captures engineering judgment, that such formality
ceilings are arbitrary gatekeeping, or that periodic human review suffices
without automation.

\textbf{Scope of this paper.} This paper does not propose a specific tool or
evaluate a system. It argues that the three properties above are necessary for
responsible AI-assisted engineering regardless of implementation. While we ground
our examples in LLM coding assistants, the framework applies equally to research
agents, planning agents, and any AI system that generates recommendations with
epistemic implications. We formalize
these properties as the First Principles Framework
(FPF)~\citep{levenchuk2023ontology,levenchuk_systems}, prove that its
aggregation semantics satisfy a quintet of invariants, and present deployment
evidence showing the practical consequences of ignoring temporal validity. A
reference implementation (anonymized for review) is an existence proof,
not the contribution itself.

% ===========================================================================
% 2. THE FIRST PRINCIPLES FRAMEWORK
% ===========================================================================
\section{The First Principles Framework}\label{sec:fpf}

This section presents FPF as a formal framework for epistemic accountability
in AI-assisted engineering. We define its core constructs, ground its
aggregation in fuzzy logic, specify the invariants any valid aggregation must
satisfy, and compare it to existing approaches.

% ---------------------------------------------------------------------------
\subsection{The F-G-R Trust Tuple}\label{sec:fgr}

Every knowledge claim in FPF carries a three-dimensional trust descriptor:

\textbf{Formality~(F):} How rigorously the claim is expressed, on a scale from
F0 to~F3.

\begin{table}[ht]
\caption{Formality levels and their reliability ceilings.}
\label{tab:formality}
\centering
\small
\begin{tabular}{@{}llp{2.8cm}r@{}}
\toprule
\textbf{Level} & \textbf{Description} & \textbf{Example} & \textbf{Cap} \\
\midrule
F0 & Informal & ``Felt faster in staging'' & 70\,\% \\
F1 & Structured & ADR with trade-offs & 85\,\% \\
F2 & Empirical & Load test: 12k~RPS & 95\,\% \\
F3 & Formal proof & TLA+, Z3 & 100\,\% \\
\bottomrule
\end{tabular}
\end{table}

The ceiling matters most: no amount of evidence can push reliability
above what the formality level permits. A decision backed entirely by informal
observations~(F0) cannot exceed 70\% reliability even if ten people agree.
This prevents informal consensus from masquerading as empirical certainty.

\textbf{Scope~(G):} Where the claim applies, expressed as a hybrid path and
tag set. Format: \texttt{path [tag1, tag2]}. Examples:
\texttt{api/auth [production, critical]}, \texttt{cache/redis [api/users]},
\texttt{*}~(universal). Scope constrains evidence transfer. A benchmark run on
a developer laptop (scope: \texttt{perf [dev, x86]}) transfers poorly to
production ARM servers (scope: \texttt{perf [prod, arm64]}). FPF formalizes
this through Congruence Levels~(CL):

\begin{table}[ht]
\caption{Congruence levels for cross-context evidence transfer.}
\label{tab:congruence}
\centering
\small
\begin{tabular}{@{}clc@{}}
\toprule
\textbf{CL} & \textbf{Context Match} & \textbf{Penalty} \\
\midrule
CL3 & Same context (internal test on target HW) & None \\
CL2 & Similar context (same arch, different scale) & 10\% \\
CL1 & Different context (external benchmark) & 40\% \\
CL0 & No context match (unrelated domain) & 90\% \\
\bottomrule
\end{tabular}
\end{table}

\noindent The penalty is applied as subtraction with a zero floor: $R_{\text{adj}} = \max(0, R(e) - \text{penalty})$. For example, CL1 evidence with $R = 0.8$ contributes $\max(0, 0.8 - 0.4) = 0.4$ to $R_{\text{eff}}$.

\textbf{Reliability~(R):} Evidence strength on $[0.0, 1.0]$, computed via
aggregation (Section~\ref{sec:min_agg_default}). $R$ is never estimated by humans; it is
always calculated from evidence scores, formality ceilings, layer ceilings
(L0\,$\leq$\,35\%, L1\,$\leq$\,75\%, L2\,$\leq$\,100\%), and dependency
structure. When both layer and formality ceilings apply, $R_{\text{eff}}$ is bounded
by the minimum of both: an F0~claim at~L1 is capped at $\min(0.70, 0.75) = 0.70$.

% Required packages for all figures (add to main paper preamble):
%   \usepackage{tikz}
%   \usetikzlibrary{arrows.meta, positioning, shapes.geometric, calc,
%                    decorations.pathreplacing, patterns, fit}
%   \usepackage{pgfplots}
%   \pgfplotsset{compat=1.18}
%   % xcolor is already loaded by icml2026.sty

% --- Color definitions used across all figures ---
%   \definecolor{fpfblue}{RGB}{70,130,180}
%   \definecolor{fpfgray}{RGB}{120,120,130}
%   \definecolor{fpfamber}{RGB}{200,150,50}
%   \definecolor{fpfgreen}{RGB}{80,160,80}
%   \definecolor{fpfred}{RGB}{190,70,70}
%   \definecolor{fpflightblue}{RGB}{210,230,250}
%   \definecolor{fpflightgreen}{RGB}{215,240,215}
%   \definecolor{fpflightamber}{RGB}{255,240,210}
%   \definecolor{fpflightred}{RGB}{250,220,220}

\begin{figure}[t]
\centering
\resizebox{\columnwidth}{!}{%
\begin{tikzpicture}[
    node distance=2.2cm,
    box/.style={
        draw=fpfblue!80!black, thick, rounded corners=4pt,
        minimum width=2.4cm, minimum height=0.9cm,
        font=\small\bfseries, text=fpfblue!90!black,
        fill=fpflightblue
    },
    outputbox/.style={
        draw=fpfgray!80!black, thick, rounded corners=4pt,
        minimum width=2.0cm, minimum height=0.7cm,
        font=\small\bfseries, text=fpfgray!90!black,
        fill=fpfgray!10
    },
    edgelabel/.style={
        font=\footnotesize, text=fpfgray!90!black, fill=white,
        inner sep=2pt, outer sep=1pt
    },
    arr/.style={-{Stealth[length=5pt]}, thick, fpfgray!70!black}
]
    % Nodes in triangle layout
    \node[box] (F) {F (Formality)};
    \node[box, below left=1.8cm and 1.2cm of F] (G) {G (Scope)};
    \node[box, below right=1.8cm and 1.2cm of F] (R) {R (Reliability)};

    % Edge: F -> R (caps ceiling)
    \draw[arr] (F) -- node[edgelabel, sloped, above] {caps ceiling} (R);

    % Edge: G -> R (constrains transfer)
    \draw[arr] (G) -- node[edgelabel, below, align=center]
        {constrains transfer\\[-1pt]{\scriptsize(CL penalty)}} (R);

    % Output arrow from R
    \node[outputbox, right=1.4cm of R] (Reff) {$R_{\mathrm{eff}}$};
    \draw[arr, fpfblue!70!black, line width=1.2pt] (R) -- (Reff);

    % Subtle annotation
    \node[font=\scriptsize, text=fpfgray, align=center, below=0.3cm of G.south west, anchor=north west]
        {Every knowledge claim carries\\all three dimensions.};
\end{tikzpicture}%
}% end resizebox
\caption{The F-G-R trust tuple. Each knowledge claim in FPF carries three
    dimensions: Formality~(F) determines the rigor of expression,
    Scope~(G) constrains evidence portability via congruence-level
    penalties, and Reliability~(R) is the computed effective trust score
    $R_{\mathrm{eff}}$. F caps the maximum achievable~R; G penalizes
    cross-context transfer.}
\label{fig:fgr-tuple}
\end{figure}
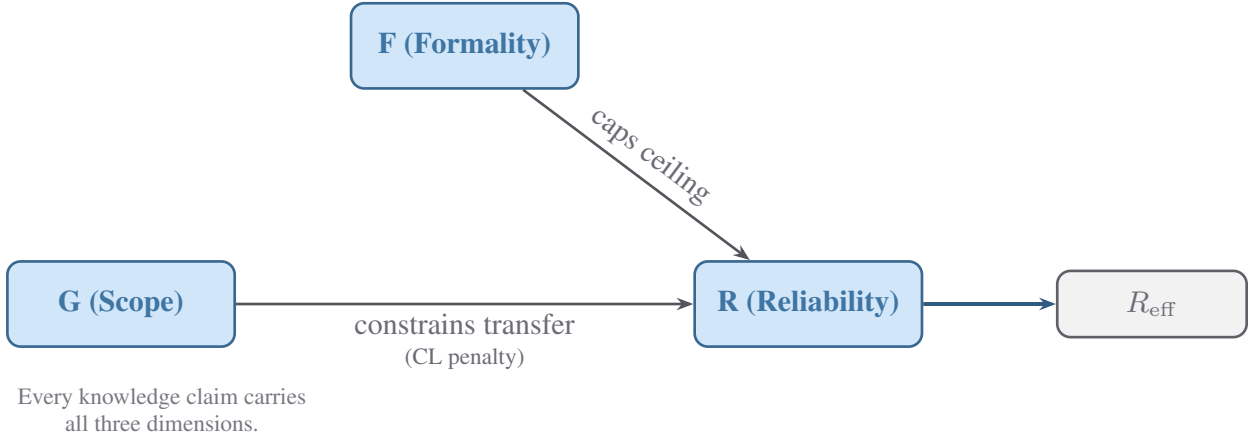

% ---------------------------------------------------------------------------
\subsection{The Gamma Invariant Quintet}\label{sec:gamma_invariant_quintet}

FPF's min-based aggregation is governed by five invariants that ensure conservative,
auditable epistemic status. These invariants apply to serial dependency
structures where argument chains must be evaluated as weakest-link systems:

\begin{enumerate}
\item \textbf{IDEM (Identity):} $\Gamma([x]) = x$. A single piece of
  evidence speaks for itself.
\item \textbf{COMM (Commutativity):} $\Gamma([a, b]) = \Gamma([b, a])$.
  Evidence order is irrelevant.
\item \textbf{LOC (Locality):} Changing evidence~$E$ does not affect holons
  with no dependency on~$E$.
\item \textbf{WLNK (Weakest Link Upper Bound):} $\Gamma(S) \leq \min(S)$.
  No aggregation may exceed the weakest link.
\item \textbf{MONO (Monotonicity):} $a \leq a'$ implies
  $\Gamma([a, b]) \leq \Gamma([a', b])$. Improving evidence never worsens
  assurance.
\end{enumerate}

\begin{theorem}[Quintet Satisfaction]\label{thm:quintet}
The G\"{o}del t-norm $\Gamma(S) = \min(S)$ satisfies all five invariants.
\end{theorem}

\begin{proof}
\textbf{IDEM:} $\min([x]) = x$ by definition of minimum over a singleton.
\textbf{COMM:} $\min(a, b) = \min(b, a)$ because minimum is symmetric.
\textbf{LOC:} $\min$ computes its result solely from its input set, with no external dependencies.
\textbf{WLNK:} $\min(S) \leq \min(S)$ trivially, so the weakest link bound is never exceeded.
\textbf{MONO:} If $a \leq a'$, then $\min(a, b) \leq \min(a', b)$ because the minimum function is monotonically non-decreasing in each argument.
\end{proof}

\begin{theorem}[Idempotent Uniqueness]\label{thm:idempotent}
The G\"{o}del t-norm is the unique idempotent t-norm~\citep{metcalfe2005fuzzy}.
\end{theorem}

\begin{proof}
Let $\ast$ be any idempotent t-norm. For $x \leq y$: by idempotence $x = x \ast x$; by monotonicity $x \ast x \leq x \ast y$; by the identity property $x \ast y \leq x \ast 1 = x$. Thus $x \ast y = x = \min(x, y)$.
\end{proof}

This uniqueness matters for practice: if we require idempotent aggregation (applying the same evidence twice should not change the result), then $\min$ is the only mathematically valid choice. Any alternative that satisfies idempotence necessarily reduces to $\min$.

The quintet is intentionally strict. Invariant~4 (WLNK upper bound) means
that even alternative aggregation functions (Product, OWA, Dempster--Shafer)
must never exceed the weakest link. These invariants prevent \emph{trust
inflation}---the failure mode where an agent hallucinates higher confidence by
aggregating massive amounts of low-quality evidence. Without this constraint, an LLM could generate ten vague observations and arithmetically combine them into apparent certainty (the epistemic equivalent of printing money).
\citet{zhang2026agentic} independently confirm this in agentic systems: the
lowest-confidence step in a reasoning trajectory predicts failure better than
any global average. Without the WLNK upper bound, three blog posts plus an
LLM summary could arithmetically exceed one controlled experiment. This is a design choice: we believe engineering
assurance should be conservative, and any relaxation requires explicit
justification. Section~\ref{sec:learnable} discusses when and how to relax
this constraint.

\begin{table*}[t]
\caption{Alternative aggregation functions and the $\Gamma$ invariant quintet.}
\label{tab:quintet-alternatives}
\centering
\small
\begin{tabular}{@{}lcccccp{5.5cm}@{}}
\toprule
\textbf{Function} & \textbf{IDEM} & \textbf{COMM} & \textbf{LOC} & \textbf{WLNK} & \textbf{MONO} & \textbf{Notes} \\
\midrule
min (G\"{o}del)     & \checkmark & \checkmark & \checkmark & \checkmark & \checkmark & Current FPF default \\
\addlinespace
Product (Lusser)    & \checkmark & \checkmark & \checkmark & $\sim$ & \checkmark & $\prod(0.9, 0.9) = 0.81 < \min = 0.9$. Satisfies WLNK when all scores $< 1$. \\
\addlinespace
OWA~\citep{yager1988owa} & \checkmark & $\sim$ & \checkmark & $\sim$ & $\sim$ & Depends on weight vector. $\mathbf{w}\!=\![0,\ldots,0,1]$ recovers min. \\
\addlinespace
Dempster--Shafer    & --- & \checkmark & \checkmark & --- & --- & Handles conflict but can exceed min. \\
\bottomrule
\end{tabular}
\end{table*}

An open research question: can we learn an optimal $\Gamma$ from historical
decision outcomes while satisfying the quintet? This requires a benchmark
dataset of architectural decisions with known outcomes
(Section~\ref{sec:learnable}).

% ---------------------------------------------------------------------------
\subsection{Min-Based Aggregation (WLNK): An Invariant-Compliant Default}\label{sec:min_agg_default}

FPF's aggregation rule is conservative: assurance equals the weakest supporting
evidence. When a decision depends on multiple pieces of evidence, the effective
reliability equals the weakest link:
\begin{equation}\label{eq:reff}
  R_{\text{eff}} = \min(\text{evidence\_scores})
\end{equation}

More precisely, when ceilings and cross-context penalties are included:
\begin{equation}\label{eq:reff-complete}
  R_{\text{eff}} = \min\bigl(
    \min_i R_{\text{adj}}(e_i),\;
    \min_j \bigl(R_{\text{eff}}(d_j) - \text{CL}_j\bigr),\;
    C_L,\;
    C_F
  \bigr)
\end{equation}
where $R_{\text{adj}}(e_i)$ is the adjusted score for evidence~$i$ (including decay, Section~\ref{sec:decay}), $\text{CL}_j$ is the congruence level penalty for dependency~$j$ (Table~\ref{tab:congruence}), $C_L$ is the layer ceiling (L0: 0.35, L1: 0.75, L2: 1.0), and $C_F$ is the formality ceiling (F0: 0.70, F1: 0.85, F2: 0.95, F3: 1.0 per Table~\ref{tab:formality}).

This is the G\"{o}del t-norm from fuzzy logic~\citep{hajek1998fuzzy}:
\begin{equation}\label{eq:godel}
  T_{\text{G\"{o}del}}(a, b) = \min(a, b)
\end{equation}

The weakest link principle has been studied in formal
argumentation~\citep{chen2023weakestlink} and critiqued for potential
over-conservatism~\citep{hoepman_weakestlink}. The G\"{o}del t-norm has four
properties that make it correct for serial argument chains:

\begin{enumerate}
\item \textbf{Commutativity:} $\min(a, b) = \min(b, a)$. The order of
  evidence does not matter.
\item \textbf{Associativity:} $\min(a, \min(b, c)) = \min(\min(a, b), c)$.
  Chaining is well-defined. (Note: associativity is a property of the specific $\min$ operator, not a required invariant for all valid aggregation functions---the $\Gamma$ quintet permits future learned aggregators that satisfy WLNK without requiring associativity.)
\item \textbf{Monotonicity:} $a \leq a'$ implies $\min(a, b) \leq \min(a', b)$.
  Stronger evidence never hurts.
\item \textbf{Boundary:} $\min(1, a) = a$ and $\min(0, a) = 0$. Perfect
  evidence is transparent; disproof is absolute.
\end{enumerate}

\textbf{Connection to possibilistic logic.}
The weakest link principle has independent theoretical grounding in possibilistic logic~\citep{dubois2025possibilistic}, where it is known as \emph{weakest link resolution}: ``the strength of an inference chain is that of the least certain formula involved in this chain.'' Possibilistic logic propagates certainty qualitatively using this law and remains inconsistency-tolerant by reasoning from the largest consistent subset of most certain formulas. FPF's WLNK bound is thus not an arbitrary design choice but a recognized principle in uncertainty reasoning with four decades of theoretical development.

\textbf{Why min, not mean?}  Consider a decision with three supporting pieces
of evidence scored at 0.95, 0.90, and~0.30. The mean is~0.72, suggesting
reasonable confidence. But the 0.30 evidence is a blog post that contradicts
published benchmarks. The $\min$~(0.30) correctly flags this: the decision
rests on a weak foundation. Averaging hides the weakness.

This matters especially for AI-assisted engineering. LLM coding assistants
produce recommendations that lack epistemic qualification, backed by training
data of varying quality. Without conservative aggregation, a team can
accumulate a portfolio of ``medium-confidence'' claims that collectively appear
strong while individually resting on weak evidence. Min-based aggregation prevents this
inflation.

\textbf{Why min is the correct default for engineering argument chains.}
Engineering decisions typically form serial dependency structures: choosing a
framework constrains library choices, which constrain API design, which
constrain test strategy. Each step depends on prior ones. A concrete comparison
illustrates the stakes. Suppose a Redis caching decision rests on three
premises: (1)~a benchmark showing adequate throughput ($R\!=\!0.95$, F2),
(2)~a traffic model predicting peak load ($R\!=\!0.70$, F1, based on a blog
post extrapolation), and (3)~vendor documentation on clustering limits
($R\!=\!0.90$, F1). Three aggregation strategies yield different results:

\begin{table}[ht]
\caption{Aggregation comparison for a Redis caching decision.}
\label{tab:aggregation-comparison}
\centering
\small
\begin{tabular}{@{}lcl@{}}
\toprule
\textbf{Aggregation} & \textbf{Result} & \textbf{Interpretation} \\
\midrule
Mean        & 0.85 & Hides weak traffic model \\
Product     & 0.60 & Penalizes more evidence \\
Min (WLNK)  & 0.70 & Identifies what to fix \\
\bottomrule
\end{tabular}
\end{table}

The mean~(0.85) suggests the decision is solid. But if the traffic model is
wrong---if peak load is 3$\times$ the blog post estimate---the entire caching
architecture fails regardless of how good the benchmark and documentation are.
Min-based aggregation surfaces this: the decision is exactly as reliable as its weakest
premise. More importantly, min-based aggregation provides actionable guidance: upgrading the
traffic model from F1 (blog extrapolation) to F2 (load test on production
traffic) would raise $R_{\text{eff}}$ from~0.70 to
$\min(0.95, 0.90) = 0.90$. No other aggregation function makes this
remediation path as transparent.

\begin{figure}[t]
\centering
\resizebox{\columnwidth}{!}{%
\begin{tikzpicture}[
    node distance=1.5cm and 0.4cm,
    drrnode/.style={
        draw=fpfblue!70!black, thick, rounded corners=4pt,
        minimum width=4.2cm, minimum height=1.0cm,
        font=\small, fill=fpflightblue, text=fpfblue!90!black,
        align=center
    },
    evidnode/.style={
        draw=#1!60!black, thick, rounded corners=4pt,
        minimum width=3.3cm, minimum height=0.9cm,
        font=\footnotesize, fill=#1!12, text=#1!90!black,
        align=center
    },
    arr/.style={-{Stealth[length=5pt]}, thick, fpfgray!60!black},
    scorelabel/.style={font=\scriptsize\bfseries, text=#1!80!black}
]
    % DRR node at top
    \node[drrnode] (drr) {
        \textbf{Use Redis for session storage}\\[1pt]
        {\footnotesize DRR \quad $R_{\mathrm{eff}} = 0.70$}
    };

    % Evidence nodes below (traffic-light: green=strong, red=weak)
    \node[evidnode=fpfgreen, below left=1.6cm and 0.8cm of drr] (E1) {
        \textbf{E1:} Benchmark\\[1pt]
        12k RPS at p95=8ms
    };
    \node[evidnode=fpfred, below=1.6cm of drr] (E2) {
        \textbf{E2:} Traffic model\\[1pt]
        peak 10k RPS
    };
    \node[evidnode=fpfgreen, below right=1.6cm and 0.8cm of drr] (E3) {
        \textbf{E3:} Redis clustering\\[1pt]
        documentation
    };

    % Edges
    \draw[arr, fpfgreen!60!black] (E1) -- (drr);
    \draw[arr, fpfred!80!black, line width=1.4pt] (E2) -- (drr);
    \draw[arr, fpfgreen!60!black] (E3) -- (drr);

    % Score labels below evidence nodes
    \node[scorelabel=fpfgreen, below=0.15cm of E1]
        {$R = 0.95$, F2};
    \node[scorelabel=fpfred, below=0.15cm of E2]
        {$R = 0.70$, F1 {\tiny(weakest)}};
    \node[scorelabel=fpfgreen, below=0.15cm of E3]
        {$R = 0.90$, F1};

    % Highlight E2 as weakest link
    \draw[fpfred!80!black, thick, dashed, rounded corners=3pt]
        ($(E2.south west)+(-0.15,-0.45)$) rectangle ($(E2.north east)+(0.15,0.1)$);

    % WLNK equation annotation
    \node[font=\scriptsize, text=fpfgray!80!black, align=left,
        below=1.6cm of E2, anchor=north] (eq) {
        $R_{\mathrm{eff}} = \min(0.95,\; 0.70,\; 0.90) = 0.70$
    };

\end{tikzpicture}%
}% end resizebox
\caption{WLNK dependency graph (worked example). Green nodes have strong
    evidence; the red node (E2, F1-level blog evidence) caps the entire
    decision at $R_{\mathrm{eff}} = 0.70$. Upgrading E2 to an F2 load
    test would raise $R_{\mathrm{eff}}$ to $\min(0.95, 0.90) = 0.90$.
    Averaging would yield 0.85, masking the weak foundation.}
\label{fig:wlnk-example}
\end{figure}

\textbf{When WLNK is correct and when it is not:}

\begin{table}[ht]
\caption{Dependency structures and WLNK applicability.}
\label{tab:wlnk-applicability}
\centering
\small
\begin{tabular}{@{}p{2.1cm}cp{3.8cm}@{}}
\toprule
\textbf{Structure} & \textbf{Valid?} & \textbf{Reasoning} \\
\midrule
Serial ($A \!\Rightarrow\! B \!\Rightarrow\! C$) & \checkmark & If any link breaks, the chain breaks. \\
Logical chains & \checkmark & Counterexample to any premise invalidates conclusion. \\
\addlinespace
Parallel redundant & --- & Defense-in-depth: joint failure $<$ weakest. \\
Independent evidence & --- & Bayesian: $P(H \mid E_1, E_2) > \max(P(H \mid E_i))$. \\
\bottomrule
\end{tabular}
\end{table}

FPF uses min-based aggregation as the default because most engineering argument chains are
serial: ``We chose Redis (premise~1: benchmark shows it is fast enough)
because (premise~2: our traffic model predicts 10k~RPS) and (premise~3: Redis
clustering handles that load).'' If any premise fails, the conclusion is
unsound. Section~\ref{sec:learnable} discusses extending the framework to
non-serial cases.

% ---------------------------------------------------------------------------
\subsection{The ADI Reasoning Cycle}\label{sec:adi}

FPF organizes reasoning as a cycle of three inference modes, each producing
claims at a higher epistemic layer. The progression mirrors Kahneman's
System~1/System~2 distinction~\citep{kahneman2011thinking}: abduction is fast
and intuitive (System~1), while deduction and induction impose slow,
deliberate verification (System~2). FPF makes this handoff explicit and
auditable:

\textbf{Abduction (generate hypotheses, L0):} Given an anomaly or design
question, generate candidate explanations. These are conjectures, plausible
but unverified. Example: ``Redis might be better than Memcached for our session
store because it supports persistence.''

\textbf{Deduction (verify logic, L0$\to$L1):} Check logical consistency. Does
the hypothesis contradict known constraints? Are the premises well-formed? A
hypothesis that passes deductive verification becomes L1 (substantiated).
Example: ``The Redis-persistence argument is logically consistent: our SLA
requires session recovery after crashes, and Redis AOF provides that. Memcached
does not.''

\textbf{Induction (gather evidence, L1$\to$L2):} Collect empirical evidence.
Run benchmarks, analyze logs, survey users. An L1 claim that passes empirical
validation becomes L2 (corroborated). Example: ``Load test confirms Redis~6.2
handles 12k~RPS at p95\,=\,8ms on our target hardware. The
session-persistence hypothesis is empirically validated.''

After the ADI cycle, a finalized decision is recorded as a Design Rationale
Record (DRR), an architectural decision record augmented with evidence
validity windows, dependency chains, and assurance scores.

\textbf{The Transformer Mandate.} We introduce this term for a structural constraint: the entity that finalizes a decision must be external to the generation loop. An LLM may propose hypotheses and gather supporting evidence (Abduction and Induction), but ratification requires an external verifier. Currently, this necessitates a human; in future multi-agent systems, the role could theoretically be filled by an independent verifier agent with disjoint training data, provided it satisfies FPF invariants, but the principle of separation of concerns remains absolute. This prevents a failure mode where an autonomous agent bootstraps confidence in its own recommendations by citing its own prior outputs. The mandate is architectural, not a policy preference. Enforcement mechanisms for this constraint remain an open research question (Section~\ref{sec:research}). \citet{ferrario2026epistemology} formalize this as computational reliabilism: the epistemic adequacy of AI-supported processes depends on whether the process is reliable \emph{for the task}, a judgment that requires external calibration.

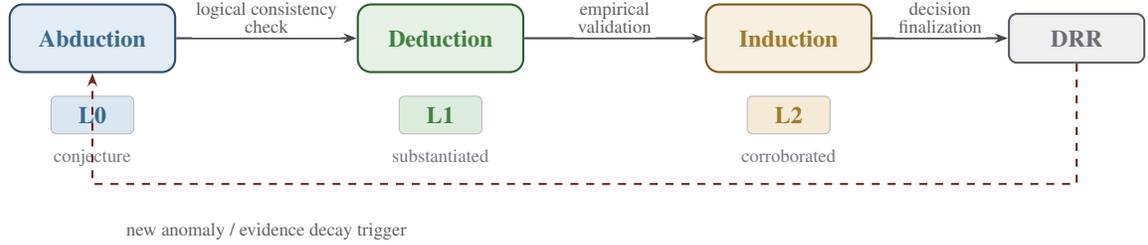
\begin{figure*}[t]
\centering
\begin{tikzpicture}[
    node distance=1.6cm and 0.6cm,
    phase/.style={
        draw=#1!60!black, thick, rounded corners=5pt,
        minimum width=2.2cm, minimum height=0.9cm,
        font=\small\bfseries, text=#1!80!black,
        fill=#1!15
    },
    layer/.style={
        draw=fpfgray!50, rounded corners=2pt,
        minimum width=1.1cm, minimum height=0.5cm,
        font=\footnotesize\bfseries, fill=#1!20, text=#1!80!black
    },
    drr/.style={
        draw=fpfgray!80!black, thick, rounded corners=3pt,
        minimum width=1.8cm, minimum height=0.65cm,
        font=\small\bfseries, fill=fpfgray!12, text=fpfgray!90!black
    },
    arr/.style={-{Stealth[length=5pt]}, thick, fpfgray!60!black},
    arrlabel/.style={font=\scriptsize, text=fpfgray!80!black, align=center, fill=white,
        inner sep=1.5pt}
]
    % Phase boxes
    \node[phase=fpfblue] (abd) {Abduction};
    \node[phase=fpfgreen, right=2.4cm of abd] (ded) {Deduction};
    \node[phase=fpfamber, right=2.4cm of ded] (ind) {Induction};

    % Layer badges below each phase
    \node[layer=fpfblue, below=0.3cm of abd] (L0) {L0};
    \node[layer=fpfgreen, below=0.3cm of ded] (L1) {L1};
    \node[layer=fpfamber, below=0.3cm of ind] (L2) {L2};

    % Layer descriptions
    \node[font=\scriptsize, text=fpfgray, below=0.08cm of L0] {conjecture};
    \node[font=\scriptsize, text=fpfgray, below=0.08cm of L1] {substantiated};
    \node[font=\scriptsize, text=fpfgray, below=0.08cm of L2] {corroborated};

    % Arrows between phases
    \draw[arr] (abd) -- node[arrlabel, above]
        {logical consistency\\[-1pt]check} (ded);
    \draw[arr] (ded) -- node[arrlabel, above]
        {empirical\\[-1pt]validation} (ind);

    % DRR box
    \node[drr, right=1.8cm of ind] (drr) {DRR};
    \draw[arr] (ind) -- node[arrlabel, above]
        {decision\\[-1pt]finalization} (drr);

    % Feedback loop: DRR back to Abduction (drop further to clear layer badges)
    \draw[arr, dashed, fpfred!60!black]
        (drr.south) -- ++(0,-1.6)
        -| (abd.south);
    % Label centered on the bottom horizontal segment
    \node[arrlabel, fill=white]
        at ($(abd.south)!0.5!(ded.south) + (0,-2.1)$)
        {new anomaly / evidence decay trigger};

\end{tikzpicture}
\caption{The ADI reasoning cycle. Abduction generates conjectures (L0),
    Deduction verifies logical consistency (L1), and Induction validates
    empirically (L2). Finalized decisions become Design Rationale Records
    (DRRs). Evidence decay or new anomalies trigger re-entry into the
    cycle.}
\label{fig:adi-cycle}
\end{figure*}

% ---------------------------------------------------------------------------
\subsection{Evidence Decay and Temporal Validity}\label{sec:decay}

Every piece of evidence in FPF carries a \texttt{valid\_until} timestamp. When
evidence expires, the system generates an alert. The team must then
re-validate the evidence, waive the decay with documented rationale, or
deprecate the decision.

This addresses a failure mode specific to AI-assisted engineering: LLMs
generate recommendations based on training data with an implicit temporal
scope. A recommendation to ``use library~X'' reflects the state of~X at
training time, not deployment time. Without temporal tracking, stale AI
recommendations persist as if they were current.

Evidence decay is not a new idea; SRE practices~\citep{beyer2016sre}
recommend regular review of operational assumptions. FPF makes it mechanical
rather than aspirational: evidence either has a validity window or it does not
count toward assurance.

\textbf{Formal definition.} Given evidence $e$ with reliability $R(e)$ and validity window $\texttt{valid\_until}(e)$, the time-dependent effective reliability is:
\begin{equation}\label{eq:decay}
  R_{\text{eff}}(e, t) =
    \begin{cases}
      R(e) & \text{if } t \leq \texttt{valid\_until}(e) \\
      0.1  & \text{otherwise}
    \end{cases}
\end{equation}

This formalization has three properties: (1)~\emph{uncertainty floor}---expired evidence moves to~0.1 regardless of its original score, representing \emph{epistemic uncertainty} rather than low confidence. When evidence expires, its conclusions---including negative conclusions---are no longer trusted. An expired falsification ($R = 0.05$) becomes ``we no longer know'' ($R = 0.1$), not ``the falsification is slightly more reliable.'' This uniform uncertainty floor ensures expired evidence is neither trusted nor distrusted, only marked as needing re-validation. This distinguishes epistemic decay from simple TTL caching; (2)~\emph{composability}---stale evidence propagates via WLNK, so any decision depending on expired evidence becomes unreliable; (3)~\emph{actionability}---the alert identifies exactly which evidence expired and which decisions are affected, enabling targeted re-validation.

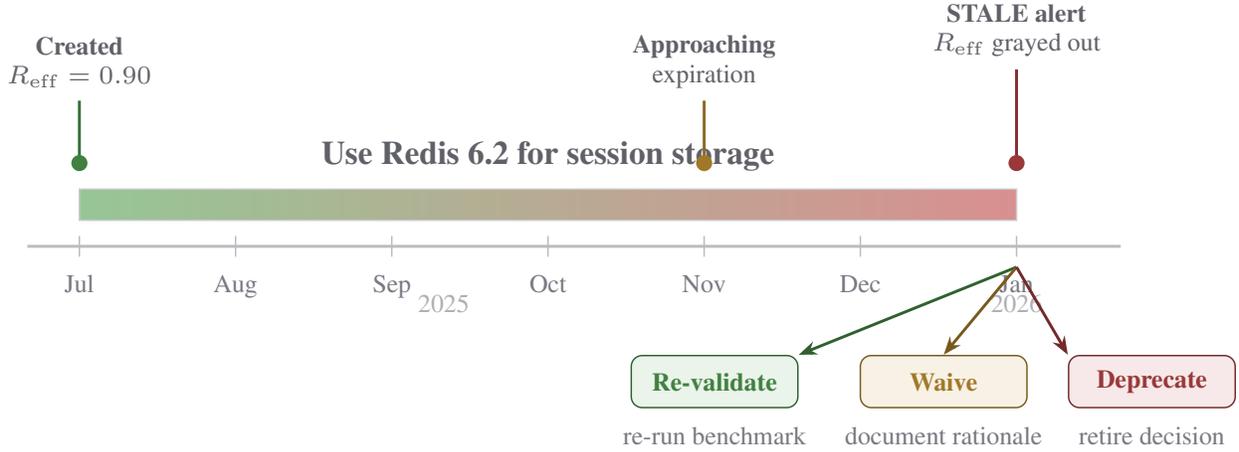
\begin{figure}[t]
\centering
\resizebox{\columnwidth}{!}{%
\begin{tikzpicture}[
    arr/.style={-{Stealth[length=5pt]}, thick, fpfgray!60!black},
    eventlabel/.style={font=\scriptsize, align=center, text=fpfgray!80!black},
    resbox/.style={
        draw=#1!60!black, rounded corners=3pt,
        minimum width=1.6cm, minimum height=0.5cm,
        font=\scriptsize\bfseries, fill=#1!12, text=#1!80!black
    }
]
    % Timeline axis
    \draw[thick, fpfgray!50] (0,0) -- (10.5,0);

    % Month tick marks and labels
    \foreach \x/\m in {0.5/Jul, 2.0/Aug, 3.5/Sep, 5.0/Oct, 6.5/Nov,
                        8.0/Dec, 9.5/Jan} {
        \draw[fpfgray!50] (\x, -0.1) -- (\x, 0.1);
        \node[font=\scriptsize, text=fpfgray, below] at (\x, -0.15) {\m};
    }
    % Year labels
    \node[font=\scriptsize, text=fpfgray!60] at (4.0, -0.55) {2025};
    \node[font=\scriptsize, text=fpfgray!60] at (9.5, -0.55) {2026};

    % Color gradient bar (green -> amber -> red)
    \shade[left color=fpfgreen!60, middle color=fpfamber!50,
           right color=fpfred!60]
        (0.5, 0.25) rectangle (9.5, 0.55);
    \draw[fpfgray!40] (0.5, 0.25) rectangle (9.5, 0.55);

    % Decision title
    \node[font=\footnotesize\bfseries, text=fpfgray!80!black, above=0.05cm]
        at (5.0, 0.55) {Use Redis 6.2 for session storage};

    % Event 1: Created (green zone)
    \draw[thick, fpfgreen!70!black] (0.5, 0.8) -- (0.5, 1.4);
    \filldraw[fpfgreen!80!black] (0.5, 0.8) circle (2pt);
    \node[eventlabel, above] at (0.5, 1.4) {
        \textbf{Created}\\$R_{\mathrm{eff}} = 0.90$
    };

    % Event 2: Approaching expiration (amber zone)
    \draw[thick, fpfamber!70!black] (6.5, 0.8) -- (6.5, 1.4);
    \filldraw[fpfamber!80!black] (6.5, 0.8) circle (2pt);
    \node[eventlabel, above] at (6.5, 1.4) {
        \textbf{Approaching}\\expiration
    };

    % Event 3: Expired (red zone)
    \draw[thick, fpfred!70!black] (9.5, 0.8) -- (9.5, 1.7);
    \filldraw[fpfred!80!black] (9.5, 0.8) circle (2pt);
    \node[eventlabel, above] at (9.5, 1.7) {
        \textbf{STALE alert}\\$R_{\mathrm{eff}}$ grayed out
    };

    % Anchor for resolution branching
    \coordinate (branchpt) at (9.5, 0);

    % Resolution paths branching from alert
    \node[resbox=fpfgreen] (rev) at (6.6, -1.3) {Re-validate};
    \node[resbox=fpfamber] (waive) at (8.8, -1.3) {Waive};
    \node[resbox=fpfred] (dep) at (10.8, -1.3) {Deprecate};

    % Arrows from STALE point to resolutions
    \coordinate (alert) at (9.5, -0.2);
    \draw[arr, fpfgreen!60!black] (alert) -- (rev.north east);
    \draw[arr, fpfamber!60!black] (alert) -- (waive.north);
    \draw[arr, fpfred!60!black] (alert) -- (dep.north west);

    % Small labels for resolution paths
    \node[font=\scriptsize, text=fpfgray, below=0.05cm of rev] {re-run benchmark};
    \node[font=\scriptsize, text=fpfgray, below=0.05cm of waive] {document rationale};
    \node[font=\scriptsize, text=fpfgray, below=0.05cm of dep] {retire decision};

\end{tikzpicture}%
}% end resizebox
\caption{Evidence decay lifecycle for a single decision. A Redis session
    storage decision is created in July~2025 with $R_{\mathrm{eff}} =
    0.90$. As the benchmark evidence approaches its validity window, the
    system transitions from green to amber. Upon expiration in
    January~2026, a STALE alert triggers one of three resolution paths:
    re-validate, waive with rationale, or deprecate.}
\label{fig:evidence-decay}
\end{figure}

% ---------------------------------------------------------------------------
\subsection{Comparison with Existing Approaches}\label{sec:comparison}

\begin{table*}[t]
\caption{FPF compared with existing approaches across six dimensions.}
\label{tab:comparison}
\centering
\small
\begin{tabular}{@{}lp{2.0cm}p{2.0cm}p{2.0cm}p{2.0cm}p{2.0cm}@{}}
\toprule
\textbf{Dimension} & \textbf{FPF} & \textbf{Self-Consist.\ \citep{wang2023selfconsistency}}
  & \textbf{Verifier Scoring \citep{lightman2024verify}} & \textbf{ADRs} & \textbf{Instr.\ Files} \\
\midrule
Epistemic layers & Explicit (L0/L1/L2) & --- & --- & --- & --- \\
\addlinespace
Aggregation & min (WLNK) with formal guarantees & Majority voting
  & Weighted scoring & N/A & N/A \\
\addlinespace
Temporal validity & Decay with timestamps & --- & --- & --- & --- \\
\addlinespace
Audit trail & Durable DRRs with dep.\ graphs & Transient & Transient
  & Static snapshots & Static file \\
\addlinespace
Scope tracking & CL0/CL1/CL2/CL3 & --- & --- & --- & --- \\
\addlinespace
Formal invariants & $\Gamma$ quintet (proven) & --- & --- & --- & --- \\
\bottomrule
\end{tabular}
\end{table*}

% ---------------------------------------------------------------------------
\subsection{Related Work}\label{sec:related}

FPF connects several research threads.

\textbf{Technical debt and architectural erosion.}
\citet{cunningham1992wycash} coined the technical debt metaphor.
\citet{kruchten2012techdebt} extended it to architectural technical debt,
identifying ``stale design decisions'' as a primary contributor. FPF
operationalizes this insight: evidence decay tracking turns ``stale
decisions'' from a metaphor into a measurable, alertable condition.

\textbf{Knowledge management in software engineering.}
\citet{robillard2010recommendation} surveyed how development teams capture and
retrieve architectural knowledge, finding that most knowledge remains tacit or
locked in outdated documents. FPF addresses this by attaching
machine-readable validity windows and dependency chains to decisions, making
knowledge staleness detectable rather than implicit.

\textbf{Engineering epistemology.}  \citet{vincenti1990engineers} argued that
engineering knowledge has distinct categories (fundamental design concepts,
practical considerations, quantitative data) that differ in how they are
produced and validated. FPF's formality levels (F0--F3) echo this taxonomy,
mapping informal observation through formal proof.

\textbf{Uncertainty quantification in machine learning.}  The ML community has
developed extensive methods for quantifying uncertainty in model predictions:
Monte Carlo dropout~\citep{gal2016dropout}, deep
ensembles~\citep{lakshminarayanan2017ensembles}, and conformal
prediction~\citep{angelopoulos2023conformal}. These approaches address a
different level of abstraction than FPF\@. ML uncertainty quantification asks
``how confident is the model in this prediction?'' FPF asks ``how reliable is
the architectural decision that deployed this model?'' A neural network may
output well-calibrated confidence scores for individual inferences while the
decision to deploy it rests on a stale benchmark, an untested scaling
assumption, and a blog post about GPU availability. FPF and ML UQ are
complementary: model-level uncertainty is one input to decision-level
assurance, not a substitute for it.

\textbf{LLM reliability and hallucination.}  \citet{ji2023hallucination}
survey hallucination in natural language generation.
\citet{huang2024hallucination} catalog failure modes in LLM reasoning. These
findings motivate FPF's conservative aggregation: if LLM-generated
recommendations carry unknown error rates, min-based aggregation provides a
worst-case bound rather than a misleading average.

\textbf{Possibilistic logic and uncertainty reasoning.}
Possibilistic logic~\citep{dubois2025possibilistic} provides the theoretical foundation for FPF's weakest link aggregation. Developed over four decades by Dubois and Prade, possibilistic logic handles weighted classical formulas where inference follows the ``weakest link resolution'' rule: the certainty of a derived conclusion equals the minimum certainty of the formulas in the derivation chain. This principle is mathematically grounded in necessity measures and has been applied to default reasoning, belief revision, and argumentation. FPF's WLNK bound is a direct application of this principle to engineering decisions, where argument chains (requirements $\to$ design $\to$ implementation) mirror possibilistic inference chains.

\textbf{Decision tracking.}  Architectural Decision Records
(ADRs)~\citep{nygard2011adr} are the closest existing practice.
\citet{jansen2005architecture} proposed knowledge management frameworks for
software architecture. \citet{esposito2025genai} survey 47 academic studies on generative AI in software
architecture, finding that 93\% of surveyed papers report no validation of LLM-generated
architectural outputs. FPF extends ADRs with temporal validity,
dependency-aware invalidation, and computed assurance scores, directly
addressing the validation gap these surveys identify.

\textbf{Argumentation and provenance.}
\citet{dung1995acceptability} introduced abstract argumentation frameworks;
\citet{besnard2008elements} extended them with structured premises. FPF shares
the graph-of-claims structure but tracks \emph{reliability} and \emph{expiry}
rather than resolving defeat.
\citet{buneman2001provenance} formalized data provenance; the W3C PROV
model~\citep{w3c2013prov} standardized provenance interchange;
\citet{josang2007survey} survey trust propagation across networks. FPF borrows
traceability and compositional trust but adds formality, scope, and temporal
validity, attributes absent from provenance records and reputation scores.

% ===========================================================================
% 3. ALTERNATIVE VIEWS
% ===========================================================================
\section{Alternative Views}\label{sec:alternatives}

% ---------------------------------------------------------------------------
\subsection{``This Is Over-Engineering''}\label{sec:alt-overengineering}

The argument: most decisions do not need this rigor. Lightweight Architectural
Decision Records (ADRs)~\citep{nygard2011adr} are industry standard and
sufficient. FPF adds bureaucracy for marginal benefit. Surveys of
practitioners~\citep{robillard2010recommendation} consistently find that teams
prefer lightweight documentation.

We partly agree. For trivial, easily reversible decisions (choosing a
date-formatting library, naming a config variable), FPF adds overhead without
value. Skip it. But for decisions with long-term consequences---database
selection, authentication architecture, data model design---hidden assumptions
compound over time. Our retrospective analysis (Section~\ref{sec:evidence}) found
that 20--25\% of such decisions showed stale evidence within two months. ADRs
document what was decided but not when evidence expires or what conditions
invalidate the decision. The real question is which decisions warrant the cost
of tracking epistemic status.

% ---------------------------------------------------------------------------
\subsection{``ADRs Already Solve This''}\label{sec:alt-adrs}

The argument: ADRs~\citep{nygard2011adr} are lightweight, widely adopted, and
sufficient for decision tracking. \citet{jansen2005architecture} showed that
architectural knowledge management can be structured without heavyweight
process.

ADRs are static snapshots. They record the decision and rationale at a point
in time but provide no mechanism for detecting when the rationale becomes
invalid. They do not track evidence expiration (benchmark results going stale),
assumption drift (traffic growing 10$\times$, hardware changing), or
dependency chains (if claim~$X$ is invalidated, which downstream decisions
break?). A Google SRE study~\citep{beyer2016sre} found that 60\% of production
outages trace to stale assumptions about system behavior. FPF complements ADRs
by adding what they lack: temporal validity windows and dependency-aware
invalidation.

% ---------------------------------------------------------------------------
\subsection{``Epistemic Layers Are Philosophically Naive''}
\label{sec:alt-epistemic}

The argument: the L0/L1/L2 hierarchy assumes a positivist epistemology where
claims progress linearly from conjecture to verified truth.
\citet{kuhn1962revolutions} showed that scientific knowledge does not
accumulate linearly, and \citet{feyerabend1975method} argued against rigid
methodological hierarchies. Real knowledge is messier: claims can be partially
verified, contextually true, or verified by one methodology and contradicted
by another.

This is the strongest objection. FPF's layers are indeed a simplification. An
L2 claim (empirically validated) might be validated by a load test that does
not capture production traffic patterns. The ``corroborated'' label could
create false confidence. However, the alternative (treating all claims as
equally uncertain) is worse in practice. Engineers already make implicit
epistemic distinctions (``I tested this'' vs.\ ``I think this should work'').
FPF makes those distinctions explicit and auditable. The layers are not claims
about ultimate truth; they are claims about what verification has been
performed. An L2 claim says ``we ran a test and it passed,'' not ``this is
certainly true.'' The scope field~(G) and congruence levels~(CL) provide the
contextual qualification that pure positivism lacks.

% ---------------------------------------------------------------------------
\subsection{``Formal Methods Are More Rigorous''}\label{sec:alt-formal}

The argument: if you want rigor, use TLA+~\citep{lamport2002tla} for
distributed protocol verification, Coq~\citep{coq_team} for algorithm
correctness, or Alloy~\citep{jackson2012alloy} for structural constraints.
FPF's formality scale is a pale imitation.

Formal methods and FPF operate at different levels. TLA+ can verify that a
consensus protocol satisfies safety and liveness properties. It cannot answer
``should we use Redis or Memcached given our traffic patterns, team expertise,
and operational constraints?'' FPF handles the empirical, contextual, and
trade-off dimensions where formal methods do not apply. They complement each
other: FPF's formality field~(F) includes formal verification as F3 evidence.
A decision backed by a TLA+ proof~(F3), a Jepsen test~(F2), and a blog
post~(F1) has $R_{\text{eff}} = \min(1.0, 0.95, 0.7) = 0.7$. The weakest
link is the blog post. The fix: replace it with a controlled experiment, not
abandon the framework.

% ---------------------------------------------------------------------------
\subsection{``WLNK Is Too Conservative''}\label{sec:alt-wlnk}

The argument: min-based aggregation ignores the value of corroborating
evidence. Bayesian epistemology~\citep{bovens2003bayesian} shows that
independent confirming evidence should increase posterior probability. Three
independent studies each scoring~0.8 should yield more confidence than a single
study scoring~0.8, but min-based aggregation treats them identically.

Correct. This is a deliberate trade-off. For serial dependencies (argument
chains, prerequisite relationships), min-based aggregation is provably correct: a chain
cannot be stronger than its weakest link. For parallel or independent evidence,
min-based aggregation is overly conservative. Section~\ref{sec:learnable} addresses this
directly: we propose extending FPF with configurable aggregation that
auto-detects dependency topology and selects the appropriate function while
maintaining the $\Gamma$ invariant quintet. The current implementation uses
min-based aggregation everywhere as a safe default. We prefer false conservatism to false
confidence.

% ===========================================================================
% 4. DEPLOYMENT EVIDENCE
% ===========================================================================
\section{Deployment Evidence}\label{sec:evidence}

This section presents evidence that architectural decision staleness is a real
and measurable problem, not a theoretical concern. We performed a retrospective
audit of two internal projects (anonymized) that used traditional ADRs without
temporal tracking. We analyzed git history and commit metadata from December~2025
to January~2026, applying FPF staleness criteria to 62 architectural decisions
to determine: (1)~which decisions had stale evidence by FPF standards, and
(2)~whether that staleness was discovered proactively or only reactively during
incidents. The question is not whether our prototype is optimal, but whether
evidence staleness occurs at rates that warrant systematic tracking. Full
methodology and per-project breakdowns are in Appendix~\ref{app:results}.

\emph{Terminology:} Evidence is \emph{stale} when the current date exceeds its validity window, the timestamp by which supporting data should be re-verified. For example, a load test result with a 60-day validity window becomes stale after 60 days, regardless of whether the underlying decision remains correct.

\textbf{AI amplification.} AI-assisted engineering amplifies the staleness problem in three ways. First, LLMs generate decisions faster than teams can validate them, increasing the rate at which potentially stale recommendations enter codebases. Second, AI recommendations carry implicit temporal scope from training data: a suggestion to ``use library~X'' reflects the state of~X at training time, not deployment time. Third, AI-generated code lacks the institutional memory that would flag dormant assumptions; an LLM cannot know that a caching decision from six months ago rested on traffic assumptions that have since doubled. These factors make mechanical validity tracking necessary.

% ---------------------------------------------------------------------------
\subsection{Evidence Decay in Practice}\label{sec:decay-results}

\begin{table}[ht]
\caption{Evidence decay metrics (retrospective audit of two projects, 2~months).}
\label{tab:decay-metrics}
\centering
\small
\begin{tabular}{@{}lr@{}}
\toprule
\textbf{Metric} & \textbf{Value} \\
\midrule
Total decisions audited & 62 ADRs \\
Stale evidence detected & 14 ($\sim$23\%) \\
\addlinespace
Discovered reactively (during incidents) & 12 (19\%) \\
Never discovered (dormant until audit) & 2 (3\%) \\
\addlinespace
Average evidence validity window & 53 days \\
Avg.\ time to debug stale decision & 4.2 hours \\
\bottomrule
\end{tabular}
\end{table}

The 20--25\% staleness rate is the central finding: nearly one in four
architectural decisions had evidence that expired within two months, regardless
of the tracking mechanism used. In projects without temporal tracking, these
stale assumptions were discovered reactively (during incidents or
refactoring), suggesting that the problem persists silently until it causes
failures.

\emph{Scope note:} This audit focused on temporal validity, which is directly
measurable from timestamps and validity windows, an O($n$) scan over $n$
decisions. Systematic measurement of formality inflation requires classifying
each decision's epistemic status (O($n \times k$) where $k$ is classification
effort), and detecting citation circularity requires tracing dependency graphs
(O($n \times m$) where $m$ is average citation depth). These higher-cost
measurements require instrumentation not present in these projects and remain
future work.

The cumulative staleness curve (Appendix, Figure~\ref{fig:staleness-curve})
shows the temporal distribution: short-validity evidence (30-day data-quality
checks) decays earliest, creating an early-warning signal.

% ---------------------------------------------------------------------------
\subsection{Proactive vs.\ Reactive Discovery}\label{sec:adr-comparison}

\begin{table}[ht]
\caption{Staleness discovery: traditional ADRs vs.\ FPF criteria.}
\label{tab:adr-comparison}
\centering
\small
\begin{tabular}{@{}lrr@{}}
\toprule
\textbf{Discovery Mode} & \textbf{Count} & \textbf{\% of Stale} \\
\midrule
Reactive (during incidents/refactoring) & 12 & 86\% \\
Never discovered (dormant until audit) & 2 & 14\% \\
Would be proactive with FPF decay alerts & 14 & 100\% \\
\bottomrule
\end{tabular}
\end{table}

The key finding: of the 14 decisions with stale evidence, 12 were discovered
only when they caused problems, during incident investigation or refactoring.
Two remained dormant, their staleness undetected until our retrospective audit.
With FPF's decay tracking, all 14 would have triggered proactive alerts before
causing incidents. The average time to understand a stale decision during
reactive discovery was 4.2~hours. We estimate that FPF's structured DRRs could
reduce this substantially by preserving decision context and evidence provenance,
though this claim requires prospective validation.

% ---------------------------------------------------------------------------
\subsection{External Evidence: The Cost of Stale Assumptions}\label{sec:external-evidence}

Our internal audit shows staleness rates; external incidents show what staleness costs. The 2012 Knight Capital failure provides a case study. On August~1, 2012, Knight Capital lost \$440~million in 45~minutes when a deployment error activated dormant code from 2003. The ``Power Peg'' market-making algorithm had been unused for nine years, but its code remained in the system. A new deployment repurposed a flag that the dormant code still monitored, causing the system to execute trades at unfavorable prices.

The root cause was not the deployment error itself but the absence of validity tracking for the decision to retain dormant code. No record existed documenting why Power Peg remained in the codebase, under what conditions it should be reviewed, or what the RNDP flag controlled. Nine years of implicit ``this code is safe to ignore'' accumulated without review. This incident illustrates the type of temporal validity gap that FPF is designed to address: a decision to retain dormant code persisted without review for nine years. While we cannot claim FPF would have prevented this specific failure, the pattern---decisions made years earlier causing failures when conditions change---matches FPF's target failure mode.

This pattern---decisions made years earlier causing failures when conditions change---is not unique to Knight Capital. \citet{ernst2015measure} surveyed 1,831 practitioners across three organizations and found that ``architectural choices are the greatest source of technical debt'' and that ``architectural issues are difficult to deal with, since they were often caused many years previously.'' The temporal distance between decision and consequence makes architectural staleness particularly insidious: by the time the failure occurs, the original rationale is lost.

% ---------------------------------------------------------------------------
\subsection{Methodological Scope}\label{sec:limitations}

This study's scope constrains the strength of its conclusions. First, the
authors built the framework and applied it retrospectively to their own
projects, a self-study design that risks confirmation bias in how evidence was
categorized and validity windows were assigned. Second, staleness rates are
sensitive to validity window calibration: shorter windows mechanically produce
higher staleness percentages, and we lack ground truth on when decisions
actually became invalid versus when their evidence formally expired. Third, the
``would have been caught proactively'' claim is counterfactual; we cannot
directly observe what FPF would have done, only simulate it. These findings
demonstrate that evidence staleness is a real and measurable problem in
engineering practice, not that FPF is the optimal solution. Stronger evidence
would require prospective deployment across multiple teams with randomized
assignment and calibration of validity windows against actual decision
invalidity events.

% ---------------------------------------------------------------------------
\subsection{Property-Based Verification}\label{sec:property-tests}

Property-based testing~\citep{claessen2000quickcheck} approximates universal quantification over input spaces by generating diverse, randomly sampled test cases. Where exhaustive verification is infeasible, PBT provides probabilistic confidence that a property holds by exercising it against thousands of automatically generated inputs, including edge cases that manual test authoring typically misses.

Five key $R_{\text{eff}}$ properties verified via property-based testing (10{,}000 iterations each): bounds $[0,1]$, WLNK enforcement ($R_{\text{eff}} \leq \min(\text{evidence})$),
formality ceiling, layer ceiling, and monotonicity. These properties test the \emph{practical correctness} of the assurance calculator, complementing the theoretical Gamma quintet (Section~\ref{sec:gamma_invariant_quintet}). All passed. Fuzz testing
(50{,}000 iterations) found zero panics on IEEE~754 edge cases (NaN, Inf,
subnormal floats). Full results in Appendix~\ref{app:results}.

% ===========================================================================
% 5. RESEARCH DIRECTIONS
% ===========================================================================
\section{Research Directions}\label{sec:research}

% ---------------------------------------------------------------------------
\subsection{Optimizing Aggregation Function Selection}\label{sec:learnable}

The First Principles Framework, through the Gamma invariant quintet, provides a
principled basis for selecting aggregation functions appropriate to specific
dependency topologies, as outlined in Section~\ref{sec:gamma_invariant_quintet}.
While min-based aggregation is the provably conservative default for serial
chains, the framework is designed to accommodate alternative functions for
non-serial evidence where Invariant~4 (the weakest link upper bound) may be
appropriately relaxed while preserving Invariants~1--3 and~5. The research
direction lies not in whether the framework can support these alternatives, but
in how to optimally select and learn aggregation functions under various
conditions.

Candidate aggregation functions for non-serial cases:

\begin{table}[ht]
\caption{Candidate aggregation functions by dependency type.}
\label{tab:learnable}
\centering
\small
\begin{tabular}{@{}lll@{}}
\toprule
\textbf{Dependency} & \textbf{Aggregation} & \textbf{Use Case} \\
\midrule
Serial ($A\!\to\!B\!\to\!C$) & min (WLNK) & Argument chains \\
Parallel (indep.) & Product & Indep.\ evidence \\
Redundant (2-of-3) & OWA & Defense-in-depth \\
Voting (multi-agent) & Self-consist. & Multi-LLM \\
\bottomrule
\end{tabular}
\end{table}

Evaluation: inter-rater reliability (automatic topology detection vs.\ expert
judgment, Cohen's $\kappa$), A/B testing (adaptive aggregation vs.\ pure min-based aggregation
on decision outcome quality), and formal verification that the learned
$\Gamma$ satisfies all five invariants.

% ---------------------------------------------------------------------------
\subsection{Federated Evidence Sharing}\label{sec:federated}

Organizations repeatedly test identical hypotheses. (``Has anyone benchmarked
Postgres~16 on ARM64?'') A federated registry where projects share benchmarks
with reproducibility metadata could reduce duplicated effort. Trust transfers
through congruence levels: CL3 (same org, same infra), CL2 (similar context),
CL1 (public benchmark). Open problems include reproducibility verification,
privacy (differential privacy on benchmark results), and adversarial evidence
(gaming assurance scores).

% ---------------------------------------------------------------------------
\subsection{SMT-Based Claim Validation}\label{sec:smt}

Quantitative claims (``API handles 1000~RPS with p95 $< 100$\,ms at
$< 80\%$ CPU'') can be expressed as SMT constraints and checked mechanically.
A SAT result constitutes F3 (formal) evidence. An UNSAT result identifies
logical inconsistency. Research question: which engineering claims are
SMT-verifiable, and which require empirical testing?

% ===========================================================================
% 6. CALL TO ACTION
% ===========================================================================
\section{Call to Action}\label{sec:calltoaction}

\textbf{For ML researchers:} Design learnable aggregation operators satisfying
the $\Gamma$ quintet that outperform WLNK on non-serial evidence. This
requires benchmarks of decisions with ground-truth outcomes, and crucially,
\emph{benchmarks that penalize epistemic drift}, where models are evaluated not
only on answer quality but on whether they detect when their own evidence has
gone stale. Study adversarial robustness of assurance scores and integration of
epistemic tracking with LLM reasoning chains.

\textbf{For practitioners:} Add \texttt{valid\_until} timestamps to
consequential ADRs and measure how often evidence expires unnoticed. Integrate
decay checks into CI/CD so benchmarks auto-refresh on dependency updates. We invite teams to apply these criteria to their existing ADR repositories and report staleness rates; independent replication would strengthen or challenge our 20--25\% finding.

\textbf{For tool builders:} Surface decision lineage and evidence freshness in
IDEs and CI/CD pipelines. Build epistemic tracking that composes with LLM
assistants through open protocols (e.g., MCP) so evidence freshness travels
with AI-generated recommendations.

The absence of epistemic accountability benchmarks is itself a research gap.
Current LLM evaluations measure output quality but not whether the model knows
when its knowledge is stale or its confidence is inflated. We call for
benchmark suites that test temporal awareness, evidence staleness detection,
and resistance to trust inflation.

% ===========================================================================
% 7. CONCLUSION
% ===========================================================================
\section{Conclusion}\label{sec:conclusion}

AI-assisted software engineering generates decisions faster than organizations
can validate them. We have argued for three properties that any responsible
AI-assisted engineering workflow should implement: explicit epistemic layers
that distinguish conjecture from verified knowledge, conservative assurance
aggregation grounded in the G\"{o}del t-norm, and temporal accountability
through evidence decay tracking.

We formalized these properties as the First Principles Framework, proved that
its aggregation satisfies a quintet of invariants, and presented deployment
evidence: \textbf{23\% of architectural decisions had stale evidence within two months},
with 86\% of that staleness discovered only during incidents. With FPF's decay
tracking, all would have been flagged proactively.

The gap between AI-generated recommendations and validated engineering
decisions will widen as LLM capabilities increase. Epistemic accountability
infrastructure is coming either way; the community can build it deliberately
or discover the need through production failures.

% ===========================================================================
% REFERENCES
% ===========================================================================
\bibliography{references}

@inproceedings{wang2023selfconsistency,
  author    = {Xuezhi Wang and Jason Wei and Dale Schuurmans and Quoc Le
               and Ed Chi and Sharan Narang and Aakanksha Chowdhery
               and Denny Zhou},
  title     = {Self-Consistency Improves Chain of Thought Reasoning
               in Language Models},
  booktitle = {Proceedings of the 11th International Conference on
               Learning Representations (ICLR)},
  year      = {2023},
  url       = {https://arxiv.org/abs/2203.11171},
}

@inproceedings{lightman2024verify,
  author    = {Hunter Lightman and Vineet Kosaraju and Yura Burda
               and Harri Edwards and Bowen Baker and Teddy Lee
               and Jan Leike and John Schulman and Ilya Sutskever
               and Karl Cobbe},
  title     = {Let's Verify Step by Step},
  booktitle = {Proceedings of the 12th International Conference on
               Learning Representations (ICLR)},
  year      = {2024},
  url       = {https://arxiv.org/abs/2305.20050},
}

@inproceedings{wei2022chainofthought,
  author    = {Jason Wei and Xuezhi Wang and Dale Schuurmans
               and Maarten Bosma and Brian Ichter and Fei Xia
               and Ed Chi and Quoc Le and Denny Zhou},
  title     = {Chain-of-Thought Prompting Elicits Reasoning
               in Large Language Models},
  booktitle = {Advances in Neural Information Processing Systems 35
               (NeurIPS)},
  year      = {2022},
  url       = {https://arxiv.org/abs/2201.11903},
}

@article{levenchuk2023ontology,
  author  = {Anatoly Levenchuk},
  title   = {Toward an Ontology for Third Generation Systems Thinking},
  journal = {arXiv preprint arXiv:2310.11524},
  year    = {2023},
  url     = {https://arxiv.org/abs/2310.11524},
}

@inproceedings{chen2023weakestlink,
  author    = {Chen Chen and Pere Pardo and Leon van der Torre
               and Liuwen Yu},
  title     = {Weakest Link in Formal Argumentation: Lookahead
               and Principle-Based Analysis},
  booktitle = {Computational Logic in Argumentation and Reasoning
               (CLAR 2023)},
  publisher = {Springer},
  year      = {2023},
  doi       = {10.1007/978-3-031-40875-5_5},
}

@techreport{hoepman_weakestlink,
  author      = {Jaap-Henk Hoepman},
  title       = {The Weakest Link Fallacy},
  institution = {Radboud University},
  year        = {2008},
  url         = {https://www.cs.ru.nl/~jhh/publications/weakest-link-fallacy.html},
  note        = {Technical Report, accessed 2026},
}

@book{hajek1998fuzzy,
  author    = {Petr H{\'a}jek},
  title     = {Metamathematics of Fuzzy Logic},
  series    = {Trends in Logic},
  volume    = {4},
  publisher = {Kluwer Academic Publishers},
  year      = {1998},
  isbn      = {978-1-4020-0370-7},
  doi       = {10.1007/978-94-011-5300-3},
}

@article{yager1988owa,
  author  = {Ronald R. Yager},
  title   = {On Ordered Weighted Averaging Aggregation Operators
             in Multicriteria Decision Making},
  journal = {IEEE Transactions on Systems, Man, and Cybernetics},
  volume  = {18},
  number  = {1},
  pages   = {183--190},
  year    = {1988},
  doi     = {10.1109/21.87068},
}

@book{beyer2016sre,
  author    = {Betsy Beyer and Chris Jones and Jennifer Petoff
               and Niall Richard Murphy},
  title     = {Site Reliability Engineering: How {Google} Runs
               Production Systems},
  publisher = {O'Reilly Media},
  year      = {2016},
  isbn      = {978-1491929124},
  url       = {https://sre.google/sre-book/},
}

@book{bovens2003bayesian,
  author    = {Luc Bovens and Stephan Hartmann},
  title     = {Bayesian Epistemology},
  publisher = {Oxford University Press},
  year      = {2003},
  isbn      = {978-0199269754},
  doi       = {10.1093/0199269750.001.0001},
}

@misc{levenchuk_systems,
  author = {Anatoly Levenchuk},
  title  = {First Principle Framework},
  year   = {2023},
  url    = {https://github.com/ailev/FPF},
  note   = {GitHub repository},
}

@article{esposito2025genai,
  author  = {Matteo Esposito and Xiaozhou Li and Sergio Moreschini
             and Noman Ahmad and Tomas Cerny and Karthik Vaidhyanathan
             and Valentina Lenarduzzi and Davide Taibi},
  title   = {Generative {AI} for Software Architecture: Applications,
             Challenges, and Future Directions},
  journal = {arXiv preprint arXiv:2503.13310},
  year    = {2025},
  url     = {https://arxiv.org/abs/2503.13310},
}

@inproceedings{cunningham1992wycash,
  author    = {Ward Cunningham},
  title     = {The {WyCash} Portfolio Management System},
  booktitle = {OOPSLA '92 Experience Report, Addendum to the Proceedings},
  publisher = {ACM},
  year      = {1992},
  doi       = {10.1145/157709.157715},
}

@article{kruchten2012techdebt,
  author  = {Philippe Kruchten and Robert L. Nord and Ipek Ozkaya},
  title   = {Technical Debt: From Metaphor to Theory and Practice},
  journal = {IEEE Software},
  volume  = {29},
  number  = {6},
  pages   = {18--21},
  year    = {2012},
  doi     = {10.1109/MS.2012.167},
}

@article{robillard2010recommendation,
  author  = {Martin P. Robillard and Robert J. Walker
             and Thomas Zimmermann},
  title   = {Recommendation Systems for Software Engineering},
  journal = {IEEE Software},
  volume  = {27},
  number  = {4},
  pages   = {80--86},
  year    = {2010},
  doi     = {10.1109/MS.2009.161},
}

@book{vincenti1990engineers,
  author    = {Walter G. Vincenti},
  title     = {What Engineers Know and How They Know It: Analytical
               Studies from Aeronautical History},
  publisher = {Johns Hopkins University Press},
  year      = {1990},
  isbn      = {978-0801845888},
}

@article{ji2023hallucination,
  author  = {Ziwei Ji and Nayeon Lee and Rita Frieske and Tiezheng Yu
             and Dan Su and Yan Xu and Etsuko Ishii and Ye Jin Bang
             and Andrea Madotto and Pascale Fung},
  title   = {Survey of Hallucination in Natural Language Generation},
  journal = {ACM Computing Surveys},
  volume  = {55},
  number  = {12},
  pages   = {1--38},
  year    = {2023},
  doi     = {10.1145/3571730},
}

@article{huang2024hallucination,
  author  = {Lei Huang and Weijiang Yu and Weitao Ma and Weihong Zhong
             and Zhangyin Feng and Haotian Wang and Qianglong Chen
             and Weihua Peng and Xiaocheng Feng and Bing Qin
             and Ting Liu},
  title   = {A Survey on Hallucination in Large Language Models:
             Principles, Taxonomy, Challenges, and Open Questions},
  journal = {arXiv preprint arXiv:2311.05232},
  year    = {2024},
  url     = {https://arxiv.org/abs/2311.05232},
}

@misc{nygard2011adr,
  author       = {Michael Nygard},
  title        = {Documenting Architecture Decisions},
  howpublished = {Cognitect Blog},
  month        = nov,
  year         = {2011},
  url          = {https://cognitect.com/blog/2011/11/15/documenting-architecture-decisions},
}

@inproceedings{jansen2005architecture,
  author    = {Anton Jansen and Jan Bosch},
  title     = {Software Architecture as a Set of Architectural
               Design Decisions},
  booktitle = {5th Working IEEE/IFIP Conference on Software Architecture
               (WICSA 2005)},
  pages     = {109--120},
  year      = {2005},
  doi       = {10.1109/WICSA.2005.61},
}

@book{kuhn1962revolutions,
  author    = {Thomas S. Kuhn},
  title     = {The Structure of Scientific Revolutions},
  publisher = {University of Chicago Press},
  year      = {1962},
  isbn      = {978-0226458120},
  note      = {4th edition 2012},
}

@book{feyerabend1975method,
  author    = {Paul Feyerabend},
  title     = {Against Method: Outline of an Anarchistic Theory
               of Knowledge},
  publisher = {New Left Books},
  address   = {London},
  year      = {1975},
  isbn      = {978-0860916468},
  note      = {4th edition, Verso, 2010},
}

@book{lamport2002tla,
  author    = {Leslie Lamport},
  title     = {Specifying Systems: The {TLA+} Language and Tools
               for Hardware and Software Engineers},
  publisher = {Addison-Wesley},
  year      = {2002},
  isbn      = {978-0321143068},
  url       = {https://lamport.azurewebsites.net/tla/tla.html},
}

@misc{coq_team,
  author = {{The Coq Development Team}},
  title  = {The {Coq} Proof Assistant},
  year   = {1989},
  url    = {https://coq.inria.fr/},
  note   = {First released 1989. See also: Coquand, T. and Huet, G. (1988).
            The Calculus of Constructions.
            \emph{Information and Computation}, 76(2--3):95--120},
}

@book{jackson2012alloy,
  author    = {Daniel Jackson},
  title     = {Software Abstractions: Logic, Language, and Analysis},
  edition   = {Revised},
  publisher = {MIT Press},
  year      = {2012},
  isbn      = {978-0262017152},
  url       = {https://alloytools.org/},
}

@inproceedings{gal2016dropout,
  author    = {Yarin Gal and Zoubin Ghahramani},
  title     = {Dropout as a {Bayesian} Approximation: Representing
               Model Uncertainty in Deep Learning},
  booktitle = {Proceedings of the 33rd International Conference on
               Machine Learning (ICML), PMLR 48},
  pages     = {1050--1059},
  year      = {2016},
  url       = {https://arxiv.org/abs/1506.02142},
}

@inproceedings{lakshminarayanan2017ensembles,
  author    = {Balaji Lakshminarayanan and Alexander Pritzel
               and Charles Blundell},
  title     = {Simple and Scalable Predictive Uncertainty Estimation
               using Deep Ensembles},
  booktitle = {Advances in Neural Information Processing Systems 30
               (NeurIPS)},
  year      = {2017},
  url       = {https://arxiv.org/abs/1612.01474},
}

@article{angelopoulos2023conformal,
  author  = {Anastasios N. Angelopoulos and Stephen Bates},
  title   = {Conformal Prediction: A Gentle Introduction},
  journal = {Foundations and Trends in Machine Learning},
  volume  = {16},
  number  = {4},
  pages   = {494--591},
  year    = {2023},
  doi     = {10.1561/2200000101},
  url     = {https://arxiv.org/abs/2107.07511},
}

@article{dung1995acceptability,
  author  = {Phan Minh Dung},
  title   = {On the Acceptability of Arguments and Its Fundamental Role
             in Nonmonotonic Reasoning, Logic Programming and $n$-Person
             Games},
  journal = {Artificial Intelligence},
  volume  = {77},
  number  = {2},
  pages   = {321--357},
  year    = {1995},
  doi     = {10.1016/0004-3702(94)00041-X},
}

@book{besnard2008elements,
  author    = {Philippe Besnard and Anthony Hunter},
  title     = {Elements of Argumentation},
  publisher = {MIT Press},
  year      = {2008},
  isbn      = {978-0262026437},
}

@article{buneman2001provenance,
  author  = {Peter Buneman and Sanjeev Khanna and Wang-Chiew Tan},
  title   = {Why and Where: A Characterization of Data Provenance},
  journal = {Lecture Notes in Computer Science},
  volume  = {1973},
  pages   = {316--330},
  year    = {2001},
  doi     = {10.1007/3-540-44503-X_20},
  note    = {Proceedings of ICDT 2001},
}

@techreport{w3c2013prov,
  author      = {Luc Moreau and Paolo Missier},
  title       = {{PROV-DM}: The {PROV} Data Model},
  institution = {World Wide Web Consortium (W3C)},
  type        = {Recommendation},
  year        = {2013},
  url         = {https://www.w3.org/TR/prov-dm/},
}

@article{josang2007survey,
  author  = {Audun J{\o}sang and Roslan Ismail and Colin Boyd},
  title   = {A Survey of Trust and Reputation Systems for Online
             Service Provision},
  journal = {Decision Support Systems},
  volume  = {43},
  number  = {2},
  pages   = {618--644},
  year    = {2007},
  doi     = {10.1016/j.dss.2005.05.019},
}

@article{singhal2023medpalm,
  author  = {Karan Singhal and Shekoofeh Azizi and Tao Tu and
             S. Sara Mahdavi and Jason Wei and Hyung Won Chung and
             Nathan Scales and Ajay Tanwani and Heather Cole-Lewis and
             Stephen Pfohl and Perry Payne and Martin Seneviratne and
             Paul Gamble and Chris Kelly and Abubakr Abdelrahman and
             Nathanael Sch{\"a}rli and Aakanksha Chowdhery and
             Philip Mansfield and Dina Demner-Fushman and
             Blaise Ag{\"u}era y Arcas and Dale Webster and
             Greg S. Corrado and Yossi Matias and Katherine Chou and
             Juraj Gottweis and Nenad Tomasev and Yun Liu and
             Alvin Rajkomar and Joelle Barral and Christopher Semturs
             and Alan Karthikesalingam and Vivek Natarajan},
  title   = {Large Language Models Encode Clinical Knowledge},
  journal = {Nature},
  volume  = {620},
  pages   = {172--180},
  year    = {2023},
  doi     = {10.1038/s41586-023-06291-2},
}

@article{boiko2023autonomous,
  author  = {Daniil A. Boiko and Robert MacKnight and
             Ben Kline and Gabe Gomes},
  title   = {Autonomous Chemical Research with Large Language Models},
  journal = {Nature},
  volume  = {624},
  pages   = {570--578},
  year    = {2023},
  doi     = {10.1038/s41586-023-06792-0},
}

@book{kahneman2011thinking,
  author    = {Daniel Kahneman},
  title     = {Thinking, Fast and Slow},
  publisher = {Farrar, Straus and Giroux},
  year      = {2011},
  isbn      = {978-0374275631},
}

@article{zhang2026agentic,
  author  = {Jiaxin Zhang and Caiming Xiong and Chien-Sheng Wu},
  title   = {Agentic Confidence Calibration},
  journal = {arXiv preprint arXiv:2601.15778},
  year    = {2026},
  url     = {https://arxiv.org/abs/2601.15778},
}

@article{ferrario2026epistemology,
  author  = {Andrea Ferrario and Alessandro Facchini and Juan M. Dur{\'a}n},
  title   = {Epistemology Gives a Future to Complementarity in
             Human-{AI} Interactions},
  journal = {arXiv preprint arXiv:2601.09871},
  year    = {2026},
  url     = {https://arxiv.org/abs/2601.09871},
}

@inproceedings{metcalfe2005fuzzy,
  author    = {George Metcalfe},
  title     = {Fundamentals of Fuzzy Logics},
  booktitle = {Lecture Notes, Tbilisi Summer School on Language, Logic
               and Computation},
  year      = {2005},
  url       = {https://www.logic.at/tbilisi05/Metcalfe-notes.pdf},
  note      = {Proves that the G\"{o}del t-norm (minimum) is the unique
               idempotent t-norm},
}

@inproceedings{dubois2025possibilistic,
  author    = {Didier Dubois and Henri Prade},
  title     = {40 Years of Research in Possibilistic Logic -- a Survey},
  booktitle = {Proceedings of the Thirty-Fourth International Joint
               Conference on Artificial Intelligence (IJCAI-25)},
  pages     = {10427--10435},
  year      = {2025},
  doi       = {10.24963/ijcai.2025/1158},
  note      = {Survey Track. Establishes ``weakest link resolution'' as
               fundamental principle of possibilistic inference},
}

@inproceedings{claessen2000quickcheck,
  author    = {Koen Claessen and John Hughes},
  title     = {{QuickCheck}: A Lightweight Tool for Random Testing of
               {Haskell} Programs},
  booktitle = {Proceedings of the Fifth {ACM} {SIGPLAN} International
               Conference on Functional Programming (ICFP '00)},
  pages     = {268--279},
  publisher = {ACM},
  year      = {2000},
  doi       = {10.1145/351240.351266},
}

@inproceedings{ernst2015measure,
  author    = {Neil A. Ernst and Stephany Bellomo and Ipek Ozkaya
               and Robert L. Nord and Ian Gorton},
  title     = {Measure It? Manage It? Ignore It? Software Practitioners
               and Technical Debt},
  booktitle = {Proceedings of the 10th Joint Meeting on Foundations of
               Software Engineering (ESEC/FSE 2015)},
  pages     = {50--60},
  publisher = {ACM},
  year      = {2015},
  doi       = {10.1145/2786805.2786848},
}
\bibliographystyle{unsrtnat}

% ===========================================================================
% APPENDICES (unlimited pages)
% ===========================================================================
\appendix

% ---------------------------------------------------------------------------
\section{Glossary}\label{app:glossary}

\begin{table}[ht]
\centering
\small
\begin{tabular}{@{}lp{5.8cm}@{}}
\toprule
\textbf{Term} & \textbf{Definition} \\
\midrule
L0   & Conjecture---unverified hypothesis \\
L1   & Substantiated---logically verified claim \\
L2   & Corroborated---empirically validated claim \\
\addlinespace
WLNK & Weakest Link---min-based reliability aggregation \\
F-G-R & Formality--Scope--Reliability trust tuple \\
DRR  & Design Rationale Record---decision artifact with validity windows \\
\addlinespace
MCP  & Model Context Protocol---stdio-based AI tool integration \\
ADI  & Abduction--Deduction--Induction reasoning cycle \\
$\Gamma$ & Universal aggregation operator (FPF spec) \\
OWA  & Ordered Weighted Averaging~\citep{yager1988owa} \\
CL   & Congruence Level---evidence transfer penalty across contexts \\
ADR  & Architectural Decision Record~\citep{nygard2011adr} \\
\bottomrule
\end{tabular}
\end{table}

% ---------------------------------------------------------------------------
\section{Implementation Architecture}\label{app:architecture}

The reference implementation is an MCP server with SQLite persistence and
ACID transactions.

\textbf{Database schema:} holons (knowledge units), evidence (with
\texttt{valid\_until} timestamps), relations (serial/parallel dependencies),
characteristics (quantitative properties), waivers (documented risk
acceptances).

\textbf{Core modules:}
\begin{itemize}
\item Assurance calculator: WLNK implementation with F-G-R propagation.
\item ADI cycle orchestration: propose, verify, validate, decide.
\item Evidence decay: staleness detection and alerts.
\end{itemize}

% ---------------------------------------------------------------------------
\section{Full Deployment Results}\label{app:results}

\subsection{Property-Based Test Results}\label{app:property-tests}

Configuration: time-based seed, parallelized execution.

\begin{table}[ht]
\caption{Property-based test results (10{,}000 iterations each).}
\label{tab:property-tests}
\centering
\small
\begin{tabular}{@{}llrc@{}}
\toprule
\textbf{Property} & \textbf{Invariant} & \textbf{Iter.} & \textbf{Status} \\
\midrule
P1 & $0 \leq R_{\text{eff}} \leq 1$ for all holons & 10k & \checkmark\ PASS \\
P2 & $R_{\text{eff}} \leq \min(\text{evidence})$ & 10k & \checkmark\ PASS \\
P3 & $R_{\text{eff}} \leq \text{ceiling}(F)$ & 10k & \checkmark\ PASS \\
P4 & $R_{\text{eff}} \leq \text{ceiling}(\text{layer})$ & 10k & \checkmark\ PASS \\
P5 & Adding evidence $\not\Rightarrow$ lower $R_{\text{eff}}$ & 10k & \checkmark\ PASS \\
\bottomrule
\end{tabular}
\end{table}

Fuzz testing: 50{,}000 iterations, zero panics on NaN, Inf, subnormal floats,
$-0.0$, values outside $[0,1]$, floating-point rounding.

% ---------------------------------------------------------------------------
\subsection{Per-Project Breakdown}\label{app:per-project}

\begin{table*}[t]
\caption{Per-project audit results (FPF criteria applied retrospectively to traditional ADRs).}
\label{tab:per-project}
\centering
\small
\begin{tabular}{@{}llclrcl@{}}
\toprule
\textbf{Project} & \textbf{Type} & \textbf{Team} & \textbf{ADRs} & \textbf{Stale (\%)} & \textbf{Evidence Type} & \textbf{Assigned Validity} \\
\midrule
A & Web app & 1 dev & 32 & 7 (22\%) & Load tests, API compat. & 60 days \\
B & Microservices & 2 devs & 30 & 7 (23\%) & Benchmarks, integration & 45 days \\
\midrule
\textbf{Total} & & & \textbf{62} & \textbf{14 (23\%)} & & \\
\bottomrule
\end{tabular}
\end{table*}

Of the 14 stale decisions identified, 12 (86\%) were discovered reactively
during incidents or refactoring. Two (14\%) remained dormant; their staleness
undetected until our retrospective audit applied FPF criteria.

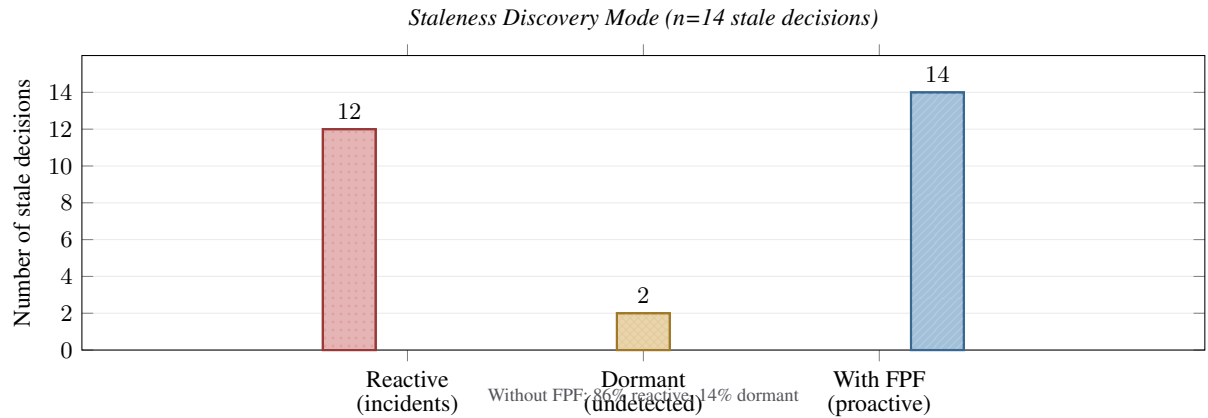
\begin{figure}[t]
\centering
\begin{tikzpicture}
\begin{axis}[
    width=\columnwidth,
    height=5.5cm,
    title={\small\textit{Staleness Discovery Mode (n=14 stale decisions)}},
    ybar,
    bar width=20pt,
    ylabel={\small Number of stale decisions},
    ymin=0, ymax=16,
    ytick={0,2,4,6,8,10,12,14},
    xtick={1,2,3},
    xticklabels={
        {\small\shortstack{Reactive\\(incidents)}},
        {\small\shortstack{Dormant\\(undetected)}},
        {\small\shortstack{With FPF\\(proactive)}}
    },
    xmin=0.3, xmax=3.7,
    enlarge x limits=0.2,
    grid=none,
    ymajorgrids=true,
    grid style={fpfgray!15},
    tick label style={font=\footnotesize},
    label style={font=\small},
    every axis plot/.append style={thick},
    nodes near coords,
    nodes near coords style={font=\footnotesize\bfseries},
    clip=false
]

    % Reactive discovery (red bar)
    \addplot[
        fill=fpfred!40,
        draw=fpfred!80!black,
        postaction={pattern=dots, pattern color=fpfred!50}
    ]
        coordinates {(1, 12)};

    % Dormant (amber bar)
    \addplot[
        fill=fpfamber!40,
        draw=fpfamber!80!black,
        postaction={pattern=crosshatch, pattern color=fpfamber!50}
    ]
        coordinates {(2, 2)};

    % Would be proactive with FPF (blue bar)
    \addplot[
        fill=fpfblue!50,
        draw=fpfblue!80!black,
        postaction={pattern=north east lines, pattern color=fpfblue!40}
    ]
        coordinates {(3, 14)};

    % Annotation
    \node[font=\scriptsize, text=fpfgray!70!black, align=center]
        at (axis cs:2, -2.5)
        {Without FPF: 86\% reactive, 14\% dormant};

\end{axis}
\end{tikzpicture}
\caption{Staleness discovery modes (retrospective analysis of 62 ADRs).
    Of the 14 decisions with stale evidence, 12 were discovered reactively
    during incidents or refactoring; 2 remained dormant until our audit.
    With FPF decay tracking, all 14 would have triggered proactive alerts.}
\label{fig:fpf-vs-adr}
\end{figure}

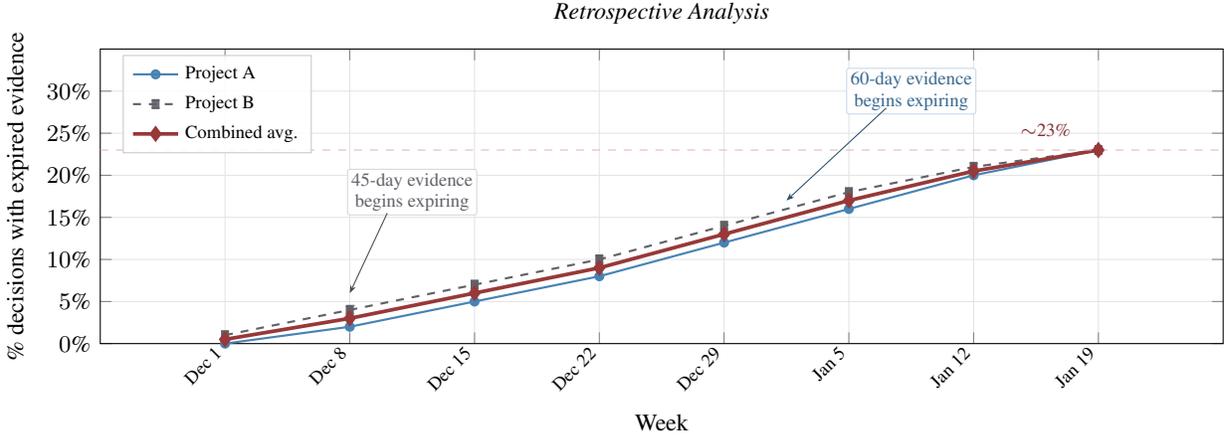
\begin{figure}[t]
\centering
\begin{tikzpicture}
\begin{axis}[
    width=\columnwidth,
    height=5.5cm,
    title={\small\textit{Retrospective Analysis}},
    xlabel={\small Week},
    ylabel={\small \% decisions with expired evidence},
    xmin=0, xmax=9,
    ymin=0, ymax=35,
    xtick={1,2,3,4,5,6,7,8},
    xticklabels={Dec 1,Dec 8,Dec 15,Dec 22,Dec 29,Jan 5,Jan 12,Jan 19},
    x tick label style={rotate=45, anchor=east, font=\scriptsize},
    ytick={0,5,10,15,20,25,30},
    yticklabel={\pgfmathprintnumber{\tick}\%},
    grid=major,
    grid style={fpfgray!20},
    tick label style={font=\footnotesize},
    label style={font=\small},
    legend style={
        at={(0.02,0.98)},
        anchor=north west,
        font=\scriptsize,
        draw=fpfgray!40,
        fill=white,
        fill opacity=0.9,
        text opacity=1,
        cells={anchor=west},
        row sep=1pt
    },
    every axis plot/.append style={thick},
    clip=false
]

    % Project A (solid line) - 60-day validity window
    \addplot[fpfblue, solid, mark=*, mark size=1.5pt]
        coordinates {
            (1,0) (2,2) (3,5) (4,8) (5,12) (6,16) (7,20) (8,23)
        };
    \addlegendentry{Project A}

    % Project B (dashed line) - 45-day validity window
    \addplot[fpfgray!80!black, dashed, mark=square*, mark size=1.5pt]
        coordinates {
            (1,1) (2,4) (3,7) (4,10) (5,14) (6,18) (7,21) (8,23)
        };
    \addlegendentry{Project B}

    % Combined average (thick, dark)
    \addplot[fpfred!80!black, solid, line width=1.4pt,
             mark=diamond*, mark size=2pt]
        coordinates {
            (1,0.5) (2,3) (3,6) (4,9) (5,13) (6,17) (7,20.5) (8,23)
        };
    \addlegendentry{Combined avg.}

    % Annotation: Project B shorter validity
    \node[font=\scriptsize, text=fpfgray!80!black, align=center,
          fill=white, inner sep=1.5pt, draw=fpfgray!30, rounded corners=1pt]
        at (axis cs:2.5,18) {
            45-day evidence\\begins expiring
        };
    \draw[-{Stealth[length=3pt]}, thin, fpfgray!60!black]
        (axis cs:2.3,15.5) -- (axis cs:2.0,6);

    % Annotation: Project A longer validity
    \node[font=\scriptsize, text=fpfblue!80!black, align=center,
          fill=white, inner sep=1.5pt, draw=fpfblue!30, rounded corners=1pt]
        at (axis cs:6.5,30) {
            60-day evidence\\begins expiring
        };
    \draw[-{Stealth[length=3pt]}, thin, fpfblue!60!black]
        (axis cs:6.3,28) -- (axis cs:5.5,17);

    % Horizontal reference line at 23%
    \draw[fpfred!40, dashed, thin]
        (axis cs:0,23) -- (axis cs:9,23);
    \node[font=\scriptsize, text=fpfred!70!black, anchor=south west]
        at (axis cs:7.3,23.5) {$\sim$23\%};

\end{axis}
\end{tikzpicture}
\caption{Cumulative evidence staleness over two months (retrospective audit).
    Applying FPF decay criteria to git history of two internal projects
    that used traditional ADRs, evidence with shorter validity windows
    (45~days in Project~B) begins decaying earlier. Both projects converge
    to $\sim$23\% stale by mid-January~2026. Without temporal tracking,
    86\% of this staleness was discovered only during incidents.}
\label{fig:staleness-curve}
\end{figure}

% ---------------------------------------------------------------------------
\subsection{Evidence Decay Examples}\label{app:decay-example}

\textbf{Example 1: Infrastructure (expiry).}
\begin{verbatim}
Decision: "Use Redis 6.2 for sessions"
Evidence: Benchmark vs Memcached 1.6
  (valid_until: 2026-01-15)
Status: EXPIRED (2026-01-22)

Alert: "Evidence expired. Actions:
  1. Re-run benchmark on Redis 7.2
  2. Waive until Q2 2026
  3. Deprecate decision"

Action: Waived until 2026-03-01
  Rationale: "Redis 7.2 upgrade Feb.
  Re-benchmark post-migration."
\end{verbatim}

\textbf{Example 2: LLM suggestion (formality).}
\begin{verbatim}
Decision: "Use FastJSON for serialize"
  Origin: LLM (Formality: F0)
Evidence: "Copilot recommended" (L0)
  + 3 Stack Overflow mentions (L0)
Status: BLOCKED - cannot reach L1

Alert: "Only L0 evidence. Needs
  benchmark or test for L1."

Action: Ran benchmark
  Result: FastJSON 40% slower.
  Deprecated, reverted to stdlib.
\end{verbatim}

\textbf{Example 3: ML model (staleness).}
\begin{verbatim}
Decision: "Deploy GPT-4-turbo"
Evidence: SWE-bench (2024-06)
  (valid_until: 2025-01-01)
Status: EXPIRED (2025-01-15)

Alert: "Benchmark predates update.
  GPT-4-turbo changed 2024-11."

Action: Re-ran on SWE-bench-Live
  Result: +12% better. Confirmed.
  Evidence refreshed.
\end{verbatim}

\textbf{Example 4: API contract (scope).}
\begin{verbatim}
Decision: "Payment API: idempotent"
Evidence: Stripe docs v2023-10
  Scope: [stripe-api, v2023-10]
Status: STALE - scope mismatch

Alert: "Scope [v2023-10] doesn't
  cover [v2024-08]. API changed?"

Action: Verified v2024-08 docs.
  Still supported. Scope updated.
\end{verbatim}

% ---------------------------------------------------------------------------
\subsection{Scalability Benchmark Details}\label{app:scalability}

Test methodology: synthetic knowledge graphs at varying sizes. Evidence
distribution: $\text{Poisson}(\lambda\!=\!5)$ per holon. Dependency structure:
40\% serial chains, 40\% parallel, 20\% isolated.

\begin{table}[ht]
\caption{Scalability benchmarks for the $R_{\text{eff}}$ calculator.}
\label{tab:scalability}
\centering
\small
\begin{tabular}{@{}rrrrr@{}}
\toprule
\textbf{Holons} & \textbf{Evidence} & \textbf{Time (ms)} & \textbf{Mem.\ (MB)} & \textbf{$\mu$s/holon} \\
\midrule
10     & 50      & 12   & 2.1  & 1\,200 \\
50     & 250     & 45   & 6.3  & 900 \\
100    & 500     & 85   & 8.4  & 850 \\
500    & 2\,500  & 210  & 28   & 420 \\
\addlinespace
1\,000 & 5\,000 & 340  & 47   & 340 \\
5\,000 & 25\,000 & 1\,850 & 180 & 370 \\
10\,000 & 50\,000 & 4\,100 & 320 & 410 \\
\bottomrule
\end{tabular}
\end{table}

Scaling: $O(n)$ time in dependency depth, $O(n + m)$ memory where
$m \approx 5n$.

\end{document}